\documentstyle[mprocl]{article}

\input{psfig}
\def\puncspace{\ifmmode\,\else{\ifcat.\C{\if.\C\else\if,\C\else\if?\C\else%
\if:\C\else\if;\C\else\if-\C\else\if)\C\else\if/\C\else\if]\C\else\if'\C%
\else\space\fi\fi\fi\fi\fi\fi\fi\fi\fi\fi}%
\else\if\empty\C\else\if\space\C\else\space\fi\fi\fi}\fi}
\def\SP{\let\\=\empty\futurelet\C\puncspace }
\def\etal{et\SP al.\SP }

\def\h-1{$h^{-1}$}
\def\lsim{~\rlap{$<$}{\lower 1.0ex\hbox{$\sim$}}}
\def\gsim{~\rlap{$>$}{\lower 1.0ex\hbox{$\sim$}}}
\def\void#1{{}}

\def\bi{\begin{itemize}}
\def\ei{\end{itemize}}

\begin{document}
\title {The ESO Imaging Survey: Status Report and Preliminary Results}
\author {L. da Costa$^1$, E. Bertin$^{1,2,3}$, E. Deul$^{1,2}$,
T. Erben$^{1,4}$. W. Freudling$^5$, M. Guarnieri$^{1,6}$, I. Hook$^1$,
R. Hook$^5$, R. Mendez$^{1,7}$, M. Nonino$^{1,8}$, L. Olsen$^{1,9}$,
I. Prandoni$^{1,10}$, A. Renzini$^1$, S. Savaglio$^1$,
M. Scodeggio$^1$, D. Silva$^1$, R. Slijkhuis$^{1,2}$, A. Wicenec$^1$,
R. Wichmann$^{1,11}$, C. Benoist$^{1,12}$}

\address{1) European Southern Observatory, Karl-Schwarzschild-Str. 2,
D--85748 Garching b. M\"unchen, Germany}

\address{2) Leiden Observatory, P.O. Box 9513, 2300 RA Leiden, The
Netherlands }

\address{3) Institut d'Astrophysique de Paris, 98bis Bd Arago, 75014 Paris,
France}

\address{4) Max-Planck Institut f\"ur Astrophysik, Postfach 1523
D-85748,  Garching b. M\"unchen, Germany}

\address{5) Space Telescope -- European Coordinating Facility,
Karl-Schwarzschild-Str. 2, D--85748 Garching b. M\"unchen, Germany}

\address{6) Osservatorio Astronomico di Pino Torinese, Strada Osservatorio 20,
I-10025 Torino, Italy}

\address{7) Cerro Tololo Inter-American Observatory, Casilla 603, La
Serena, Chile}

\address{8) Osservatorio Astronomico di Trieste, Via G.B. Tiepolo 11, I-31144
Trieste, Italy}

\address{9) Astronomisk Observatorium, Juliane Maries Vej 30, DK-2100
Copenhagen, Denmark}

\address{10) Istituto di Radioastronomia del CNR, Via Gobetti 101,
40129 Bologna, Italy}

\address{11) Landensternwarte Heidelberg-K\"onigstuhl, D-69117,
Heidelberg, Germany }

\address{12)  DAEC, Observatoire de Paris-Meudon
		5 Pl. J. Janssen
		92195 Meudon Cedex, France
 }

\maketitle

\section {Introduction}

The ESO Imaging Survey (EIS) presented in earlier issues of the
Messenger~\cite{renzini}$^,$~\cite{dacosta}, and with up-to-date
information on the ongoing observations available on the web
(http://www.eso.org/eis), is a concerted effort by ESO and the Member
State community to provide targets for the first year of operation of
the VLT. It consists of two parts: a relatively wide-angle survey
(EIS-WIDE) to cover four pre-selected patches of sky, 6 square degrees
each, spread in right ascension to search for distant clusters and
quasars and a deep, multicolor survey in four optical (SUSI-2) and two
infrared bands (SOFI) covering the HST/Hubble Deep Field South (HDFS)
and its flanking fields (EIS-DEEP).  From the start, the main challenge
has been to carry out a public survey in a limited amount of time
requiring observations, software development and data reduction with the
goal of distributing the survey data products before the call for
proposals for the VLT. To cope with this one-year timetable, a novel
type of collaboration between ESO and the community has been established
which has allowed EIS to combine the scientific and technical expertise
of the community with in-house know-how and infrastructure. In spite of
adverse weather conditions in some of the earlier runs, EIS has already
proved to be a successful experiment achieving most of its scientific
and technical goals, thereby laying the ground work for future imaging
surveys.

\section {Observations}

Observations for EIS-WIDE are being carried out with EMMI on the
NTT. They started July 1997, and so far 36 out of 42.5 half and
full-nights (360 hours) have been used, with data being accumulated in
four patches over eight runs. All observing runs have been carried
out, in standard visitor mode, by members of the EIS team. In contrast
to most earlier work, the EIS mosaic consists of frames with
significant overlaps (a quarter of an EMMI frame). The easiest way of
visualizing the geometry of the EIS mosaic is to picture two
independent sets of frames, each forming a contiguous grid (normally
referred to as odd and even), superposed and shifted in right ascension
and declination by half the length of an EMMI frame. In this way, each
position on the sky, except at the edges of the patch, is sampled by at
least two independent frames for a total integration time of 300
sec.  To ensure continuous coverage, adjacent odd/even frames have a
small overlap at the edges ($\sim 20$ arcsec). Therefore, a small
fraction of the surveyed area is covered by more than two
frames. Such a mosaic ensures good astrometry, relative photometry and
the satisfactory removal of cosmic hits and other artifacts.

EIS was one of the first programs to use the upgraded NTT and as
expected had to overcome some problems both on and off the
telescope. These included: some limitations of the current version of
the Phase-2 Proposal Preparation software (P2PP) for large programs;
unexpected overheads from the new data flow system (DFS) and VLT
control system (VCS); failures of the EMMI controller (CAMAC);
problems with the pointing model and difficulties in retrieving data
from the ESO Archive. The pointing model was particularly relevant to
EIS because of the mosaic pattern adopted. Thanks to the dedication of
people in the NTT team, the User Support Group, the Archive Group and
the EIS team these issues have been largely overcome. As expected, EIS
has proved to be a useful test case for supporting the smooth
transition between the first releases of software engineering products
and routine science operations at the NTT.

These problems and the need for calibration of the new filters have led
to some time losses. However, the impact that these time losses have
had on the overall performance of the survey has been relatively minor
as compared to the losses due to bad weather. This is illustrated in
Figure 1, which shows for each EIS run the percentage of time used for
observations. The figure does not tell the whole story as dismal
full-night runs such as runs 4 and 5 had a considerably larger impact
on the sky coverage than earlier half-night runs.  Furthermore, the
quality of nights has also varied considerably within a run and from
run to run. These difficulties were brought to the attention of the
EIS Working Group (WG) which recommended several adjustments in the
scope of the observations, such as giving priority to the I-band
observations and limiting the scope of the B-band coverage of patch
B. However, the changes will not severely affect the primary science
goals of the survey as originally proposed, with the exception of the
search for high-redshift ($ z \gsim 3$) QSOs.

So far a total of about 2000 science frames, roughly equivalent to 20
square degrees, have been taken. In Table 1, we list the position of
the centers (J2000) of the various patches and the current sky
coverage, in square degrees, for the different passbands.  In order to
maintain some degree of uniformity of contiguous regions, most regions
observed under poor conditions have been re-observed. In Figure 2, we
show the seeing distribution for all the I-band frames that have been
accepted for coaddition up to run 7. Even though these re-observations
have limited the sky coverage, some 3 square degrees have been
re-observed, they have been worthwhile since currently the median
seeing in the different patches is in the range 0.8-1.1 arcsec. We
should emphasize that since EIS is being carried out over a fixed
amount of time, a trade-off between area and data quality is
unavoidable.

\medskip
\begin{center}
 {{\bf Table 1.  Current Sky Coverage }}

\end{center}
\medskip

\begin {center}
\begin {tabular}{lcccccc}

\hline
Patch   & $\alpha$ & $\delta$& B  & V   & I  \\
\hline
A       & 22:42:54  &   -39:57:32 & -   & 1.2  & 3.2 \\
B       &  00:49:25 &   -29:35:34 & 1.5  & 1.5  & 1.6  \\
C       &  05:38:24  &  -23:51:00  &  -   &  -   & 5.6  \\
D       &   09:51:36 &   -21:00:00 &-   &  -   & 2.8  \\
\hline
        &    -       &      -      & 1.5  & 2.7  & 13.2  \\
\hline
\end{tabular}

\end{center}

\medskip

\section {EIS Pipeline}

By far the most demanding task of the EIS team has been the
development of a fully automated pipeline to handle the large data
volume generated by EIS. A major aim of the EIS software is to handle
the generic problem posed by building up a mosaic of overlapping
images, with varying characteristics, and extraction of information
from the resulting inhomogeneous coadded frames. The long-term goal
has been to develop a ``portable'' system, which may eventually be
installed in other European institutes.

Due to time constraints, it was decided early on that the development
of the pipeline should take advantage, as much as possible, of
preexisting software elements such as: 1) standard IRAF tools for the
initial processing of each input image; 2) the Leiden Observatory Data
Analysis Center (LDAC) software, developed for DENIS~\cite{denis}, to
perform photometric and astrometric calibrations; 3) the SExtractor
object detection and classification code~\cite{bertin}; 4) the
``drizzle'' image coaddition software~\cite{fruchter}$^,$~\cite{hook},
originally developed for HST, to create coadded output images from the
many, overlapping, input frames.

However, handling mosaic data taken on different nights under varying
conditions has required significant changes in the preexisting software
and the need for new concepts and intermediate products to provide the
necessary information for the source extraction and data quality control
of the coadded superimage, from which the final object catalogs are
created. In order to illustrate the power of the tool being developed a
brief description of the architecture of the pipeline is necessary.  For
each input frame a weight map, which contains information about the
noise properties of each pixel in the frame, and a flag map, which contains
information about the pixels that should be masked, such as bad pixels
and likely cosmic hits, are produced.  After background subtraction and
astrometric and relative photometry calibration, each input frame is
mapped to a flux-preserving conical equal area projection grid, chosen
to minimize distortion in area and shape of objects across the
relatively large EIS patch. The flux of each pixel of the input frame is
redistributed in the superimage and coadded according to weights of the
input frames contributing to the same region of the coadded image.

In the process of coaddition combined weight and context maps are
created. The combined weight map provides the information necessary for
the object detection algorithm to adapt the threshold of source
extraction to the noise properties of the context being analyzed.
SExtractor has been modified to incorporate this adaptive thresholding. 
The context map characterizes the origin of each pixel of the superimage
and provides information which relates each detected object to the set
of input frames that have contributed to its final flux.  A context
should be viewed as a virtual frame with its depth and seeing being
almost uniform and determined by the combination of a unique set of
input frames. For a survey such as EIS, being carried out in visitor
mode with varying seeing conditions, the context information is
essential as it may not be possible to easily characterize the PSF in
the final coadded image, which can compromise the reliability of the
galaxy/star classification algorithm.  More importantly, the contexts
represent regions of uniform noise, seeing and depth. Therefore, using
the information available for each context one may {\it a posteriori}
define ``uniform'' regions with well-defined limiting magnitudes from
which object and derived catalogs (e.g. the candidate cluster catalog)
may be extracted.

The pipeline works equally well for stacking dithered images and, in
this case, the context map can be used to easily carve out the deepest
part of the coadded image. Images from EMMI, SUSI and DENIS have been
used for tests providing excellent results.

Another area, common to all large programs, that has demanded
considerable attention is the bookkeeping and monitoring of the progress
of the observations, the data reduction and data quality. This has
required interfacing the pipeline to a database, with calls installed in
the various modules of the pipeline, which allows the status of a
particular frame or a set of frames (corresponding to ten EMMI exposures
that make up an EIS observational block) to be tracked throughout its
lifetime in the pipeline and archive. In addition, logs of the
observations and the reductions, reports (monitoring the progress of the
observations, down time, survey efficiency, etc.) and diagnostics produced
by the different modules of the pipeline (monitoring the pointing, seeing,
astrometry and relative photometry) are created and posted on the web
automatically for easy access by all team members.

In order to allow full control of the data flow, from the preparation of
the observations to the final archiving of the data, it is essential to
have a suitable data acquisition system which can be interfaced to the
data reduction pipeline, such as the DFS/VCS implemented on the NTT.
Even though some additional features are still required most of the
basic tools are already in place. There are areas where improvements can
be made such as the mass production of observational blocks (OBs),
essential for large programs with well-defined observing protocol, and
the automatic generation of the observed schedule, which would allow the
automatic updating of the status of the OBs resident in the repository.
The DFS has been a remarkable undertaking that will have an enormous
impact on the efficiency with which large programs will be conducted at
the VLT. Similar systems, even if simpler versions, should be considered
for other telescopes dedicated to survey work.

Even though considerable work lies ahead and several tests of the
performance and fine-tuning of different modules of the pipeline
remain to be carried out, it has been possible to streamline the
pipeline to only a few scripts that control the entire process of data
reduction. Under normal conditions, the data reduction requires four
commands to go from raw frames residing in the Archive to coadded
frames to object catalogs back into the Archive.  The typical data
flow rate is of 0.032~MB/sec on a 250 MHz UltraSparc2, which means
that the reduction of a 6 square degree patch requires about 70
hours. We should point out that at this stage no attempt has yet been
made to multiplex operations, and the the flow rate given above should be
considered as a very conservative lower limit.

The pipeline is being developed in close collaboration with the ESO
Archive group. The ESO Archive is the entry point of the raw data into
the EIS pipeline and the distribution point of the final products to the
community. Furthermore, we are using the EIS data to prototype general
purpose tools which will be required by the Science Archive Research
Environment. This collaboration has been mutually beneficial yielding:
1) a general server to support EIS catalogs (fits binary tables) and the
display of conical equal area coadded image sections; 2) the
installation of a server to display the superimage, stored in 4k
$\times$ 4k sections; 3) the implementation of additional features in
SKYCAT; 4) The implementation of the EIS database; 5) the development of
algorithms to cross-correlate the EIS catalogs with other available
databases.

\section {Preliminary Results}

In order to illustrate the final products of the pipeline, Figure~3
shows a low-resolution (3 arcsec/pixel) representation of the I-band
coadded image of patch A. This low-resolution image is created
automatically by the pipeline and serves as a preliminary check of the
coaddition. Also shown are the associated weight and context maps.  In
patch A, there are over 3500 contexts for 300 input frames, spanning a
wide range of scales. About 520 are "big" contexts where there are two
overlapping images. The average size of these is roughly $3.7' \times
3.7'$.  While most patches have been observed primarily in I-band, 1.5
square degrees of patch B have been covered in B, V and I, and as an
illustration Figure~4 shows a true color composite image of a small
area with multicolor information.

The final catalog of objects extracted from the coadded image uses the
weight map to adapt the threshold of the source extraction algorithm and
for each detected object identifies the context in which it has been
found. The characterization of each context is critical to control of
the completeness and uniformity of the object catalogs derived from the
coadded image. This is currently the main area of software development.

One of the major concerns has been whether the observations would be
deep enough to search for distant clusters of galaxies. To address this
and other questions posed by the WG and OPC, the EIS team has also
developed basic tools for the scientific evaluation of the data, which
allow for the comparison of star and galaxy counts with those of other
authors and for the search of clusters using the matched-filter
algorithm. A preliminary comparison of EIS galaxy counts with those
obtained by the Palomar Distant Cluster Survey (PDCS)~\cite{postman}
shows that the EIS galaxy counts extend beyond those of the PDCS. 
Preliminary tests with EIS I-band data also indicate that about 15-20
candidate clusters per square degree can been found in the estimated
redshift range $0.2 < z < 1.2$, which will yield over 300 candidate
clusters after the completion of EIS-WIDE.  Tests have also shown that
``complete'' cluster samples will be able to be produced by an adequate
selection of ``uniform'' areas within the survey region using the {\bf
context} information.  The astrometry of the EIS catalogs has an
internal accuracy of about 0.03 arcsec, more than adequate for
multi-slit spectroscopy at the VLT. The relative photometric accuracy is
estimated to be of the order of 0.05 mag. However, the absolute
calibration of patch A is still uncertain as it relies to a large extent
on data not yet available from other telescopes.

\section {Data Release: Catalogs and Images}

EIS will produce a wide range of data products including the following
catalogs:

\begin{enumerate} 

\item Single frame catalogs

\item Object catalogs extracted from the coadded images

\item Color catalogs

\item Derived catalogs (e.g. candidate clusters)

\end{enumerate}

The object catalogs will include positions, magnitude, major and
minor-axis, position angle, stellarity index, SExtractor flags, and in
the case of the catalogs extracted from the coadded image, the context
and the characteristics of the context such as noise, a measure of the
seeing, the 1$\sigma$ limiting isophote and the limiting magnitude
for point sources in that context.

The following pixel maps will also be available in the ESO Archive:
\begin{enumerate}

\item Astrometrically and photometrically calibrated frames

\item Coadded images, weight and context maps at both full resolution and
in compressed form

\item Cutouts of selected regions

\end{enumerate}

Currently, the EIS team and the ESO Archive group are studying the ways
and means of making the EIS data accessible to the community. The target
date of July 31, 1998 has been proposed to the OPC for the full release
of the data from calibrated frames to coadded images to object catalogs and
derived catalogs. The date is a compromise that takes into account the
workload of the EIS team and the call for proposals for the VLT,
expected to be issued on August 1, 1998.  Given the large amount of data
involved a questionnaire is available on the EIS home page to survey the
type of data products the astronomical community is most likely to
request in the final release.  The results of this survey will help
evaluate the demand and the definition of a general policy that
optimizes the distribution of the data to the largest possible number of
astronomers in the community.

However, recognizing the value that a preview of the data would have,
even if in a preliminary form, it has been decided that some products
will be distributed earlier. A tentative schedule for the release of
these products is available on the EIS home page
(http://www.eso.org/eis), with the first release expected to be in
March, soon after the publication of the present issue of the
Messenger. In this first release we expect to make available reduced
single frames with astrometric and photometric calibration, and
corresponding catalogs.  In addition, the compressed coadded image,
weight and context maps of patch A and the corresponding preliminary
catalog will be available on-line. This release will also be
accompanied by a complete description of the observations, data
reduction and the contents of the catalogs as well as other information
characterizing the quality of the data and catalogs, prepared by the
EIS team. A catalog of cluster candidates will also be produced for
distribution together with image cutouts to serve as finding charts.

The hope is that this preliminary release will trigger valuable
feedback from the community and will already provide useful data for
the preparation of VLT projects.  These preliminary releases will also
serve as a test case for the final delivery, allowing the EIS team and
the Archive group evaluate the amount and type of requests, the
performance of the on-line server and the requirements posed by the
distribution of large amounts of data in the form of pixel maps.
Since several applications may not require the data at full
resolution, the EIS team is currently analyzing the performance of
various compression algorithms, and their impact on the astrometry and
photometry. This information is available on the EIS home
page and, whenever possible, this option should be considered as it
would greatly facilitate the distribution of large volumes of data.

Finally, it is important to emphasize the enormous workload of the EIS
team. It includes observations, software development, tests of the
pipeline, data reduction of survey frames as well as from other
telescopes for the photometric calibration of the survey. Moreover, the
team is also involved in the preparation for EIS-DEEP and the upgrade of
the software to handle data from the wide-field imager at the 2.2m
telescope. Therefore, the team will not be able to provide support to
the users until the final release of the data. For the time being,
technical inquiries should be addressed to the ESO Archive Group
(awicenec@eso.org with copies to eisweb@eso.org). Time allowing, answers
to the most frequently asked questions will be made available on the EIS
home page.

\section {EIS Software and By-products}

Besides the astronomical data, the EIS team is committed to make
publicly available to the ESO community all the software that has been
developed for EIS. Our ultimate goal is to make, as much as possible,
the EIS pipeline portable to other institutes. By July 1998, a detailed
description of the various algorithms used in the different modules of
the EIS pipeline should become available. Full documentation and
in-depth discussion is also envisioned by the end of 1998, if the
necessary resources are available. For that purpose, an effort is
being made to keep the complete history of EIS on the web for future
reference including documentation, summaries of meetings, results of
tests, error reports, software upgrades, and relevant communication
between team members. Although time consuming, this activity has been
considered an integral part of the development phase in view of the
long term prospects of the EIS pipeline and the expected turnover of
the members of the EIS team.

In addition to these more tangible products, EIS has also provided
other important by-products including valuable information on the
performance of the NTT and the DFS/VCS. It has been used to prototype
several developments in the ESO Science Archive Research Environment
(SARE) and has created some useful new interfaces between ESO and the
community, such as the EIS WG and the EIS team.

This shows the usefulness of public surveys not only for finding
astronomical targets for the VLT but also from the operational point
of view. It has also shown the agility of ESO in responding to a
challenge, attracting talent spread throughout the ESO community and
coordinating an effort such as EIS at short notice.  Without
all of these elements it would have been impossible to keep to the
ambitious timetable of EIS.

\section {The EIS team and the Involvement of the ESO Community}

The community has played an active role in the survey by participating
in the EIS Working Group, which met three times during 1997 to decide on
the modifications of the survey strategy as they became necessary. The
ESO Member State community is also broadly represented in the EIS team,
which is composed primarily of visitors and represent over 80\% of the
4.5 FTEs allocated to the project in 1997.

Equally important has been the contribution given by the Geneva
Observatory, which has monitored the extinction measurements during
the EIS observations, and the Leiden Observatory, which has provided
time for observations of standard stars and several pointings of the
fields being observed, as well as DENIS data to be used for external
calibration. H. Boehnhardt and collaborators also supplied
observations taken at the 2.2m telescope. All of these efforts have
been extremely helpful in order to recover from the time losses caused by
El Ni\~no, providing essential data for the calibration of the
photometric zero-point of the different patches.

EIS has also captured the interest of the astronomical community with
the EIS home page being accessed about 3500 times during the month of
January 1998 and averaging 4000 hits per month in 1997.

\section {Conclusions}

Even though the weather was not very cooperative at the beginning of
the survey, it is fair to say that EIS is achieving most of its
originally planned goals. A detailed account of the ongoing work has
been presented to the EIS WG and to the OPC. As a result the OPC has
given the go ahead for the continuation of EIS-WIDE and has allocated
time for EIS-DEEP.  It should be mentioned that since SOFI will only
be available in June, the EIS WG has recommended that the goals of the
original DEEP-I and II~\cite{renzini} be combined to cover the HDFS
region and its flanking fields, with the observations scheduled to
start in July 1998.

Observations for EIS-WIDE with EMMI will be completed in March 1998, to
be followed by U-band observations with SUSI-2 in the fall of 1998 over
about 1.5 square degrees of patch B, already covered in B, V and I. At
the same time the preparations have started for the EIS-DEEP
observations. Trial reductions with single frames, taken with EMMI and
SUSI, have shown that the EIS pipeline can adequately handle dithered
optical images. Attention will now turn to interfacing EIS with the SOFI
data reduction pipeline.

The EIS pipeline is already a reality, taking raw data and producing
coadded images, object catalogs and derived catalogs, largely
un-supervised. Most of the remaining work is to implement and verify the
production of final object catalogs with the required "context"
information for data quality control, essential in the preparation of
complete samples for statistical studies. The pipeline has been
developed with one eye on short-term needs and the other on the
long-term, which should facilitate its upgrade to handle the data from
the wide-field camera at the ESO/MPIA 2.2m telescope.

Preliminary results clearly indicate that the EIS data meet the
requirements for the primary science goals of the project which, in
conjunction with the various by-products outlined above, make this pilot
program a success. One of the important remaining challenges is to make
the data reach the community in a timely and easy to use manner.
Hopefully, by doing so EIS will pave the way for gradually more
ambitious public surveys.

\bigskip

\noindent {\bf  Acknowledgments}

We would like to thank the members of the EIS Working Group, for their
scientific input in the definition of the survey, and the OPC for their
support of this pilot program.  We would also like to acknowledge the
support of DMD, in particular M. Albrecht, the NTT team, the ECF, in
particular R. Albrecht, and Jim Beletic for innumerable discussions. We
would like to express our thanks to A. Baker, D. Clements, S. Cot\'e, E.
Huizinga and J. R\"onnback for their contribution in the early phases of
the EIS project and C. Lidman for providing data from his survey. We are
also grateful to Yannick Mellier, Marc Postman and Steve Kent for their
help on several fronts and for providing some test data, and D. Koo for
valuable comments during his visit to ESO.

\newpage

\noindent {\bf Figure Caption}

\bigskip

\noindent {\bf Figure 1- } Fraction of allocated time used for
observations (filled circles) as a function of EIS run in the period
July 1997 to January 1998, compared to the average (over the past five
years) number of clear nights (upper line) and photometric nights (lower
line) per month at La Silla. Note that runs 1 and 2 were half-nights.

\medskip

\noindent {\bf Figure 2- } Seeing distribution, as measured from the
I-band images in all four EIS patches. The large-seeing tail in the
distribution of patch A are frames for which no new observations were
possible due to the lack of time.

\medskip

\noindent {\bf Figure 3- } Coadded I-band image of patch A (top panel)
showing a region approximately $2^\circ \times 1^\circ$ and the
corresponding weight (middle panel) and context maps (bottom
panel). In the weight map dark colors represent regions of higher
noise.

\medskip

\noindent {\bf Figure 4- } Color composite obtained from a selected
region $13' \times 7.5'$ of the coadded B, V and I images of patch
B. Note the presence of ghost images around bright stars present in
the B and V filters. In the region a nearby cluster ($z \sim 0.1$) is
seen as well as a concentration of red galaxies at the lower-left of
the figure.

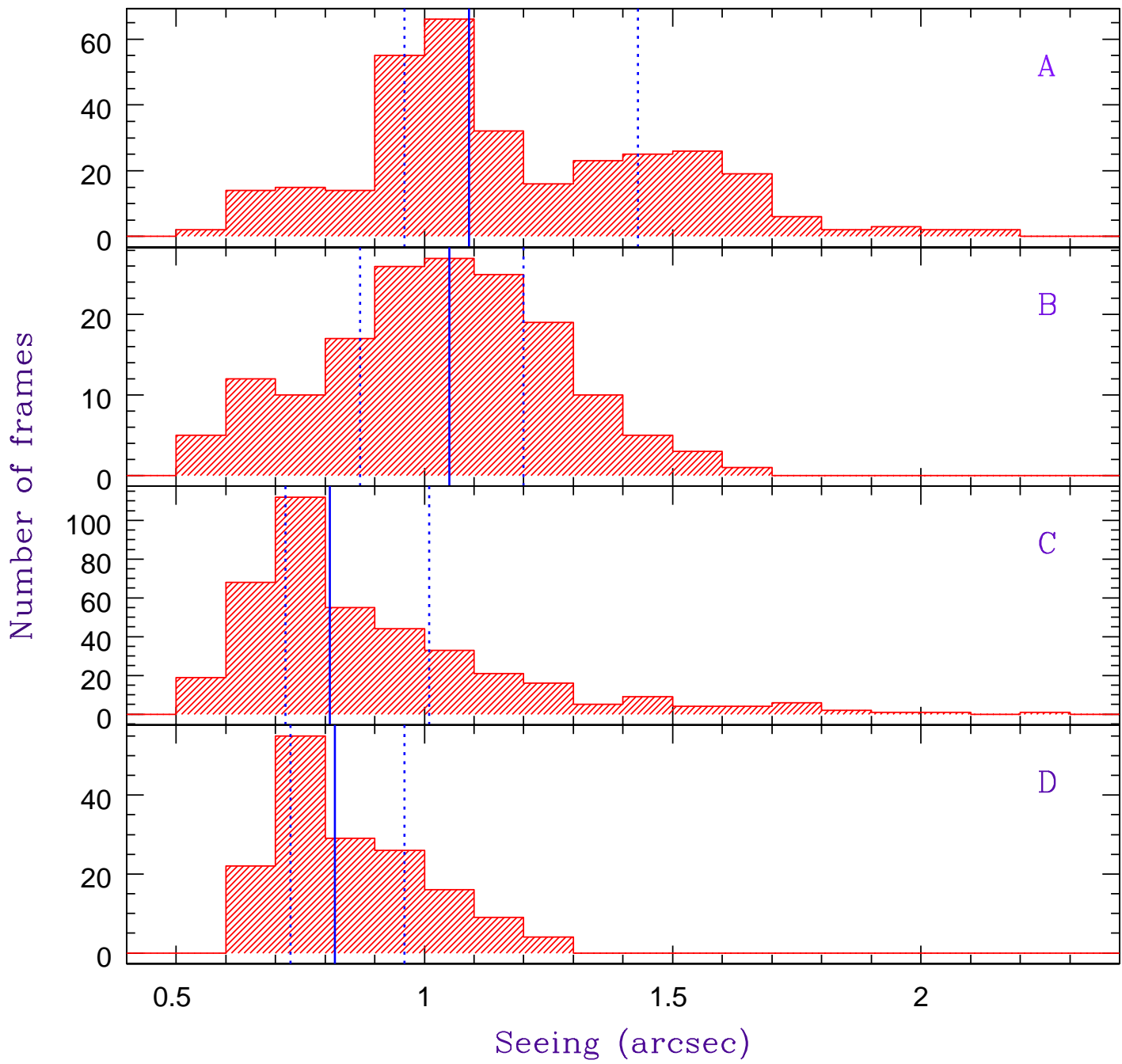
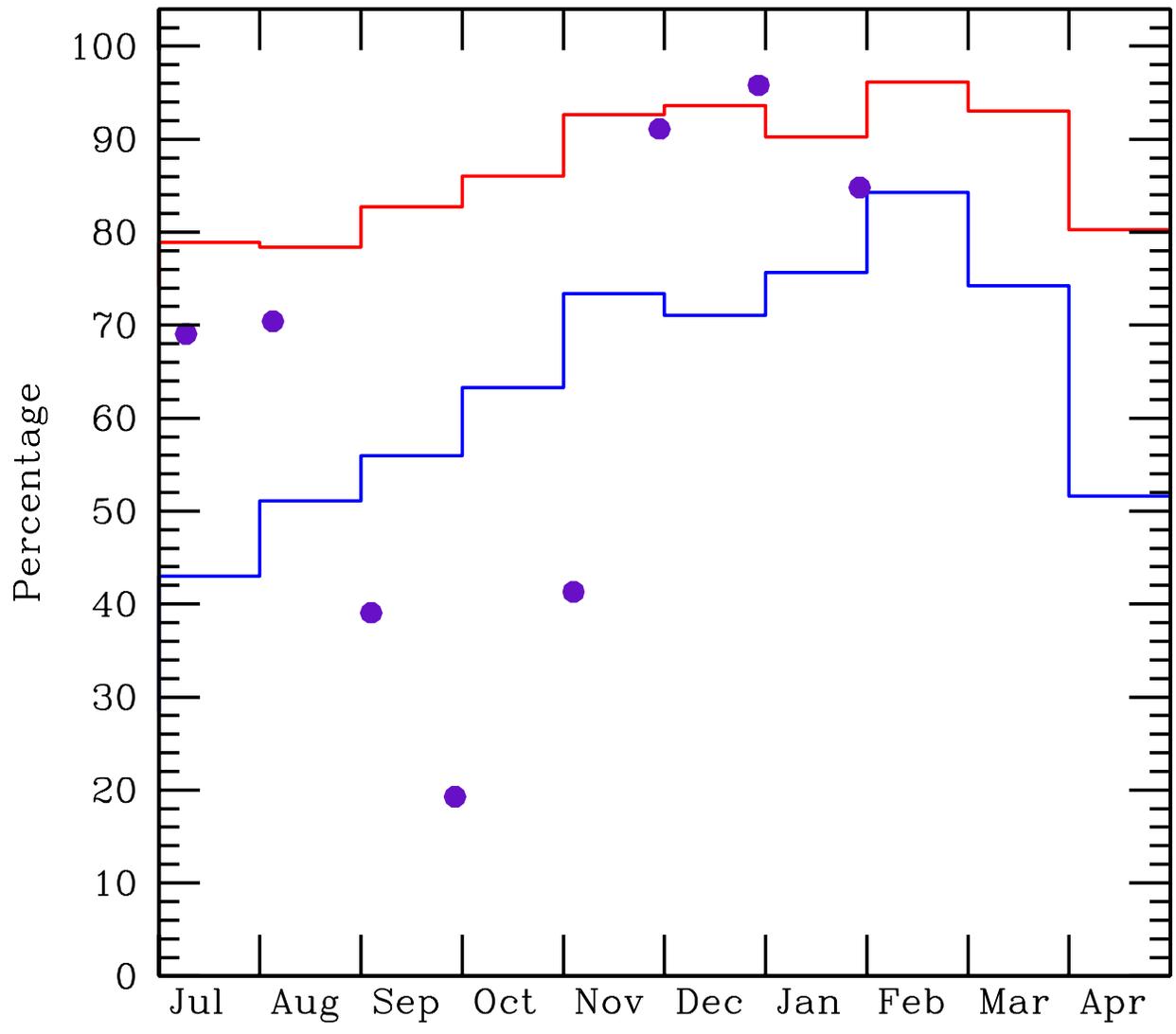
\end{document}